# 交通流演化动力学研究的一般框架及其性质分析

**摘要**：
　　针对道路网络需求时变的交通流达到均衡的条件以及均衡状态的稳定性问题，运用演化博弈理论和动力系统稳定性理论，提出了流量演化动力学研究的一般框架，建立了多群体多准则出行选择流量演化系统的动力学模型，证明了流量演化动力学模型平衡点与动态交通流量分配模型均衡解之间的等价性，讨论了模型解的存在性、唯一性和稳定性。理论分析结果表明，流量演化动力系统的演化稳定策略等价于动态交通流分配模型的均衡解，流量演化动力系统在某个局部区域总存在着唯一解，并且在出行者个体收益参数满足一定条件的前提下，流量演化动力系统在平衡点附近会表现出不同的稳定性质，通过一个简单路网的数值算例说明了本文建立模型的合理性和有效性。本文在出行者博弈行为和动态交通分配之间建立了联系，有助于加深对于城市交通流演化规律的认识和理解。

**关键词**：城市交通；交通流；动力学；交通分配；演化博弈

**中图分类号**：N945.1；C94　　　**文献标识码**：A

## 1. 引言

　　近十年来，城市交通基础设施的建设成绩令人瞩目，但交通拥堵现象不但没有好转，反而愈演愈烈。这一现象使得交通管理者和研究人员认识到：城市交通问题是一个复杂的系统工程，交通系统是一个影响因素众多、多层次相互关联、目标功能多样、非线性动态演化的开放复杂巨系统[1]，交通流是将巨大的微观离散个人决策结果转化为道路与网络上的宏观聚集现象，而交通拥挤是作为交通需求结果的交通流在特定时间空间下所表现的一种状态。要解决城市交通拥挤问题就必须结合出行者的出行决策，用数学物理模型刻画交通流量的网络分布，揭示与解释城市交通流的自组织演变规律与拥堵突现轨迹。本文关注的是网络上日复一日（day-to-day）交通流的动力学建模和演化行为，具体而言，网络上需求随时间变化的交通流能否达到均衡，在什么样的前提条件下会达到均衡？均衡的稳定性如何？这里涉及到预测型（predictive）的动态均衡问题，回答这些问题，将有助于加深我们对于城市交通流演化规律的认识和理解，对于发展出有效的交通流管理控制措施也具有重要的理论价值和实际意义。

　　由于日复一日（day-to-day）的交通流模型能够刻画拥挤网络上出行者的学习和网络达到均衡的过程，因此在交通流动力学的研究中受到较多关注，如 Horowitz[2]提出了个体路径选择的随机动力系统模型。Smith[3]考虑了驾驶员变换路径的交通分配问题的动力系统模型。Nagurney 和 Zhang[4]针对固定需求下的静态交通分配问题，为出行路径选择调整过程建立了投影动力系统。Jin[5]提出了满足先进先出（first-in-first-out，FIFO）条件的动力系统模型。Yang 和 Zhang[6]回顾了五种典型的用户调整过程，指出在满足驾驶员理性假设条件下，稳态网络流量和确定性用户均衡是等价的。不同于上述基于路径的动力学模型，He 等人[7]为避免基于路径模型的初始值依赖和路径重叠影响问题，提出了基于路段的动力学模型，给出了模型稳定性严格的数学分析，Han 和 Du[8]在 He 等人的基础上建立了非对称情形下的扩展模型，给出了关于稳定性更多的解析结果。

　　除此之外，Peeta 和 Yang[9]研究了装备先进信息系统的可操作的路径导航控制策略的稳定性问题，提出了动态交通分配问题稳定性分析的一般程序。Cho 和 Hwang[10]研究装备了

预测出行信息的车辆交通网络的日常流量演化,证明了动力系统的均衡解满足 Wardrop 均衡条件。Mounce[11]针对确定型固定需求的交通分配问题,对一日内和日间两种情况进行了动力系统建模。Cantarella 和 Cascetta[12]比较了确定情形下日常出行的动力系统模型和传统的用户均衡方法,使用非线性动力系统理论得出了不动点和均衡状态稳定性的条件。Jabari 和 Liu[13,14]针对随机情形下的交通流动力系统通常采用在确定型流量模型中加入随机项可能带来负的交通密度的问题,提出了一个新的随机动力系统模型,通过一个实际路网的数据对模型的合理性和有效性进行了验证。

从研究方法而言,现有的对于交通流演化动力学的研究主要包括动力系统理论建模、基于博弈论的仿真建模和将这两种方法结合起来进行建模等三种研究方法,上述文献大多是采用动力系统理论进行数学分析的思路。

也有一些文献从博弈论的角度出发,运用理论分析或者仿真实验,结合出行者出行行为对交通流演化过程和机理展开研究。如李振龙[15]对诱导条件下驾驶员路径选择行为进行了演化博弈分析。熊轶等人[16]在有交通信息系统作用下的路网中,建立了一个等价的随时间演化的随机用户均衡模型。刘天亮和黄海军[17,18]将日常路径选择看作是一个长期的多方非合作博弈过程,对出行信息公开和不公开两种情形下的系统演化过程进行了模拟研究。唐毓敏和冯苏苇[19]基于博弈论和 Arnott 瓶颈模型,探讨政府采用拥挤收费政策参与博弈情形下,出行者行为以及道路总成本的改变。田琼和黄海军[20]建立了基于乘客的出发时间选择的动态出行均衡模型,并对具有不同运输能力的公交方式所导致的乘客出行差异进行了比较分析。Arentze 和 Timmermans 等人[21]开发了基于强化学习和适应的动态选择模型框架,在该模型中指出了学习系统所需要的组件。Menasché 等人[22]针对出行者在交通瓶颈路段接收到多媒体交通信息以后来决定使用率的行为,提出了一个新的双层规划模型。Sun 等人[23]在交通信号灯博弈中应用遗传算法和合作演化技术开发了两种互相竞争的策略:攻击者和生存者策略,结果证明该技术可以用来开发更为复杂的策略或者合作模型。Yang 的论文[24]从演化博弈的角度对由于出行者具备学习、调整能力从而导致日常交通流从不均衡到均衡动态变化这一过程进行了深入分析。Treiber 和 Kesting[25]综合运用实证、仿真实验、理论推导的方法,研究了拥挤交通流的不稳定传递现象,通过实测数据给出了拥挤传播参数的估计,并推导出不稳定传递的分析准则。

从现有研究成果可以看到,目前对于日常交通流动力学建模和演化机理方面的研究,大多集中于确定型或者随机型的静态交通分配,本文在现有研究的基础上,讨论网络上需求随时间变化条件下交通流的演化动力学,建立动态交通分配和出行者博弈行为之间的联系。

本文在引言部分首先对交通流动力学的研究现状进行评述,然后在第二部分提出流量演化动力学研究的一般框架,第三部分建立流量演化系统的动力学模型,第四部分讨论交通流动态分配模型均衡解与流量演化动力学模型平衡点之间的等价性,第五部分讨论流量演化系统解的特性,第六部分给出了一个由两节点两条路段组成的简单路网数值算例,最后是结束语,简要总结了本文的主要内容并指出了未来进一步的研究方向。

## 2. 流量演化动力学研究的一般框架

本文提出的对于交通流量演化分析的思路如下,首先建立出行选择的流量演化模型,然后证明流量演化模型平衡点与网络动态均衡解两者之间的等价性,如果能够证明两者之间的等价关系,则可以从流量演化模型本身直接分析流量的稳定性,而不需要对动态交通分配模

型求取解析解，这一部分的工作将主要使用数学解析方法进行研究，研究得到的结果是流量演化动力学的宏观特征。下面的图 1 是本文提出的交通流演化动力学分析的一般研究框架。

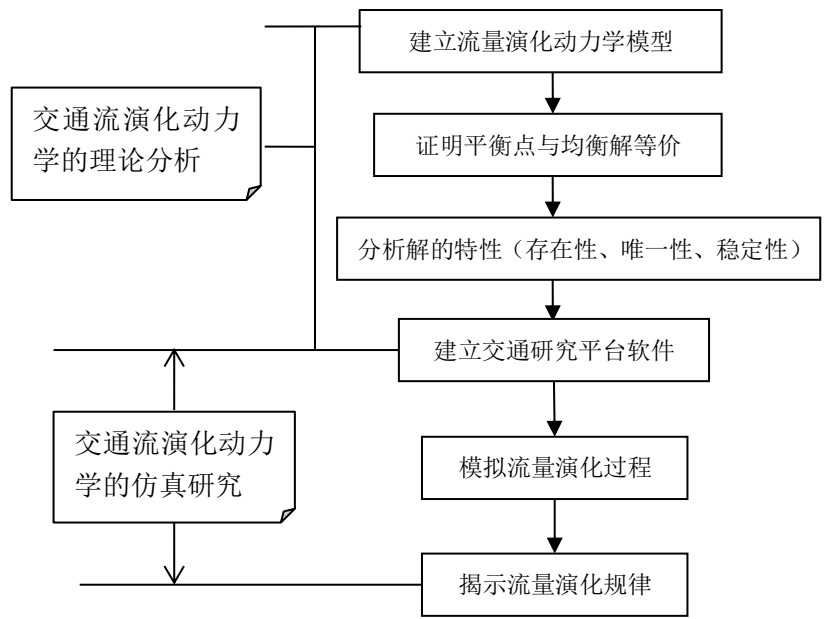

图 1 交通流演化动力学的研究框架

Figure 1　Research framework of traffic flow evolutionary dynamics

下面就图 1 所示的一般研究框架的具体含义进行说明。

第一步是建立出行选择的流量演化动力学模型，其主要步骤如下：

（1） 设定构成演化动力学模型博弈关系的局中人、策略集、支付函数和行动顺序等几个基本要素。

① 确定局中人，本文中的局中人为大量不同类别的异质出行者，具有学习和适应能力，能够根据外部环境的变化不断地调整自己的出行行为；

② 确定局中人的策略集，本文中出行者的出行策略是在出行过程中进行出行方式、出发时间和出行路径的单个或者多个选择；

③ 确定支付函数，计算局中人的支付，这一步首先要根据问题的特点，找出能正确描述效用数量关系的支付函数的具体形式，然后根据支付函数来对各种博弈情形下出行者的效用值进行计算，一般考虑的效用是按照与时间相关和时间无关两部分构成，这反映了出行个体依据不同的准则进行出行决策；

④ 确定博弈的行动顺序，根据建模的问题进行具体分析。

（2）选择动态方程，选取一个合适的动态方程来描述流量演化的动力学过程。

（3）建立微分动力系统，根据第（2）步选择的动态方程，得到表示系统流量演化过程的微分动力系统方程组。

第二步，等价性证明，即证明流量演化模型平衡点与动态交通流分配模型均衡解两者的等价性。其证明思路是，因为流量演化模型平衡点可以分成纯策略 Nash 均衡点和混合策略 Nash 均衡点两类，首先证明满足动态确定型用户最优条件（DUO）的流量等价于博弈模型中纯策略 Nash 均衡点，再证明满足动态随机型用户最优条件（DSUO）的流量等价于博弈模型中混合策略 Nash 均衡点，而动态交通流分配模型是根据这两种最优条件建立的，它的

均衡解就是满足这两种最优条件下形成的均衡流量。这一步通过流量演化模型这个中间桥梁，在动态交通流分配问题和出行者博弈行为之间建立了联系。

第三步，在第二步等价性证明的基础上，对流量演化动力学模型解的性质进行分析，包括解的存在性、唯一性性和稳定性。只有在有解的情况下，讨论解的稳定性才有意义，因此首先需要讨论演化模型解的存在和唯一性，在此基础上，研究解的各种稳定性成立的充分条件。

第四步，在前面几步工作的基础上，选择合适的开发平台（如 swarm、Repast 或者 Visual Studio 等）和编程语言（如 C\C++、Java 等）建立交通模拟研究平台软件。

第五步，利用开发的交通流研究平台软件，对交通流演化动力学进行数值模拟研究，在各种典型的算例网络上，对不同出行选择行为，如出行方式、出发时间、出行路径和多维联合出行选择等情形下的交通流演化过程进行动态模拟，考察流量演化过程和出行选择行为决策特征参数之间的关系，揭示系统在各种出行行为下流量的演化规律。

## 3. 流量演化系统的动力学模型

本小节将讨论流量演化系统动力学模型的建立，首先确定演化系统动力学模型中博弈关系的局中人（出行者）、局中人策略集、支付函数、博弈行动顺序等几个基本要素的取值，然后运用微分动力系统理论建立流量演化动力学模型。

### 3.1 演化动力学模型博弈结构设定

(1) 确定局中人。设出行总体数量为 $N$，按照某种规则（如，收入）将这 $N$ 个个体划分为 $I = \{1,2,\cdots,n\}$ 共 $n$ 种类型，因为出行者群体总共有 $n$ 种类型，这些群体之间的博弈可以看作是 $n$ 个不同类型的出行者之间进行的非对称博弈。

(2) 确定局中人的策略集。假设出行策略包含有选择交通方式、出发时间和出行路径三个基本内容，出行者可选的交通方式集为 $M = \{m_1, m_2, \cdots, m_{|M|}\}$ 共 $|M|$ 种类型，因为道路网络的基本构成单元是起讫点对，所以我们选择任意一个起讫点对，对它上面的交通流演化动态进行考察，设 $rs$ 为路网中任一起讫点对，其中 $r \in O \subset N, s \in D \subset N$，O,D 分别为起始点集合和终止点集合。出行者从起点 $r$ 出发至终点 $s$，出行者可供选择的出发时间集为 $B = \{b_1, b_2, \cdots, b_{|B|}\}$ 共 $|B|$ 个出发时间，起讫点对 $rs$ 间的路径集合为 $R = \{r_1, r_2, \cdots, r_{|R|}\}$ 共 $|R|$ 条，这里的路径是广义路径，由路段和交叉口组成。对于某种交通方式 $m_j \in M$，设其可行的行驶路径的条数为 $l_{m_j}$ 条，其可行的行驶路径为整个路径集 $R$ 上的一个子集，设为 $R_{m_j} = \{r_1, \cdots, r_{l_{m_j}} : r_p \text{为交通方式} m_j \text{可行的路径}, p \in \{1, \cdots, l_{m_j}\}\} \subseteq R$。

对于某类型的出行者 $i \in I$，设其可行的出行方式为：

$M_i = \{m_1^i, \cdots, m_{J_i}^i : m_p^i$ 为类型 $i$ 可行的出行方式, $p \in \{1, \cdots, J_i\}\} \subseteq M$，共有 $J_i$ 种。出行者 $i$ 可行的出发时间为 $B_i = \{b_1^i, \cdots, b_{K_i}^i : b_p^i$ 为类型 $i$ 可行的出发时间, $p \in \{1, \cdots, K_i\}\} \subseteq K$，共有 $K_i$ 个。出行者 $i$ 可行的出行路径为 $R_i = \bigcup R_{m_p}, p \in \{1, \cdots, J_i\} \subseteq R$。出行者 $i$ 的出行策略 $s_i$ 为出行方式、出发时间和出行路径的组合，其策略集为 $S_i = \{s_i : s_i = (m_i, b_i, r_i) : m_i \in M_i, b_i \in B_i, r_i \in R_i\}$。

(3) 确定支付函数。对于本文研究的问题而言，出行者出行产生的效用为路径经过的路段上的效用和交叉口效用之和，设 $U_i(s)$ 表示第 $i$ 个出行者采用出行策略 $s = (s_i, s_{-i})$ 下获得的效用（这里的效用 $U_i(s)$ 的值为负的，意味着出行者出行将产生支出），其表达式为：

$$U_i(s) = \theta_i \lambda \bar{U}_{it}(s) + (1 - \theta_i) \hat{U}_{it}(s) \tag{1}$$

上式中，等号右边第一项表示与时间相关的效用，第二项是与时间无关的效用，这里参数 $\theta_i \in [0,1]$ 表示不同类型出行者对于时间相关效用的重视程度，$\theta_i$ 越大表示对时间相关效用越重视，对时间无关效用越不重视，反之，$\theta_i$ 越小表示对时间相关效用越不重视，对时间无关效用越重视。参数 $\lambda > 0$ 是时间费用的转换系数，起到将时间和非时间成本的单位进行统一的作用。其中，时间相关效用的计算公式为：

$$\bar{U}_{it}(s) = \sum_{a \in R_i} \bar{U}_{it}^a(s) \tag{2}$$

上式等号右边每一项 $\bar{U}_{it}^a(s)$ 表示出行者 $i$ 采用策略 $s$ 时，在路段 $a$ 上产生的时间相关效用，由路段行走时间和交叉口延误时间两部分构成。

$\hat{U}_{it}(s) = -c_i(s)$ 是与时间无关的出行费用，其大小与交通出行费用、安全性、便利程度、舒适度、服务的弹性和可靠性有关，不影响到问题分析的情况下可以统一将 $c_i(s)$ 取为某种交通方式下的金钱支出，如果是公交，则是乘车费用，如果是小汽车，则换算成一次出行付出的费用。计算局中人的支付关键是要确定支付函数的计算公式中各项参数的值，一般可按照经验值进行选取。

(4) 确定博弈参与方的行动顺序。这里假设出行者进行的是重复的日常交通出行行为，比如说往返于两地的上下班通勤，而每天的博弈过程是 $n$ 种类型的群体出行者对出行策略的执行结果，如果我们以天为时间单位，则一天当中不同的出发时间可以看作是在同一个时间段上，所以出行者每天的出行博弈可以看作是同时执行的阶段博弈，从长期的时间尺度来看，

导致交通流动态演化的博弈可以看作一个反复进行的多人博弈过程。

### 3.2 流量演化系统动力学模型

根据演化博弈理论，建立演化动力学模型的关键在于根据问题的需要选取合适的动态方程，通常有五种形式的动态方程可供选择[26, 27]。本文假设所有的个体出行者都按照效用最大化原则进行决策，这里选择复制者动态方程建立刻画系统演化动态的动力学模型。记 $n$ 种类型的出行总体的纯策略个数分别为 $w_1, w_2, \cdots, w_n$，总的出行策略数目 $w = \sum_{i=1}^{n} w_i$，记出行总体策略数目向量 $\mathbf{w} = (w_1, w_2, \cdots, w_n)^T$，记 $N_i = \{1, \cdots, w_i\}$，$N_i$ 表示第 $i$ 个出行者采用的纯策略序号的集合，指标集 $W = \{1, \cdots, w\}$。相同群体中出行者将采用相同的出行策略，$t$ 时刻第 $i$ 个出行者采用的出行策略为 $\mathbf{x}_i(t) = (x_{i1}(t), x_{i2}(t), \cdots, x_{iw_i}(t))^T$，这里，$x_{ij}(t)$ 表示 $t$ 时刻在第 $i$ 种类型的群体中采取第 $j$ 个纯策略所占的比例，也可理解为第 $i$ 个出行者采取第 $j$ 个纯策略的概率，不失一般性，可假设其为线性函数。$t$ 时刻出行总体采用的策略为 $\mathbf{x}(t) = (\mathbf{x}_1(t), \cdots, \mathbf{x}_n(t))^T$，其中，$\mathbf{x}_i(t)$ 中的每一个分量满足概率条件 $0 \leq x_{ij}(t) \leq 1, \sum_{j=1}^{w_i} x_{ij}(t) = 1$。$t_0$ 时刻为建模的起始时刻，设 $t_0$ 时刻第 $i$ 个出行者采用的出行策略为 $\mathbf{x}_{i0} = (x_{i1,0}, \cdots, x_{iw_i,0})^T$，即 $\mathbf{x}_i(t_0) = \mathbf{x}_{i0}$，$t_0$ 时刻对整个群体满足初始条件 $\mathbf{x}(t_0) = \mathbf{x}_0$。用 $w_i$ 维单位向量 $\mathbf{e}_{i,j} = (0; \cdots, 1; \cdots, 0)^T$ 表示第 $i$ 种类型群体的第 $j$ 个纯策略，这里 $j \in \{1, \cdots, w_i\}$，$\mathbf{e}_{i,j}$ 的第 $j$ 个分量为 1，其他分量等于 0。记 $n$ 个类型的个体出行者选择选择纯策略 $(\mathbf{e}_{1,k_1}, \cdots, \mathbf{e}_{n,k_n})^T$ 时，出行者 $i$ 的收益为 $u_i(\mathbf{e}_{1,k_1}, \cdots, \mathbf{e}_{n,k_n})$，其中，$\forall i \in I, k_i \in N_i$。

根据复制者动态方程

$$\frac{\mathrm{d} x_{ij}(t)}{\mathrm{d} t} = x_{ij}(t)[h_{ij}(\mathbf{x}(t)) - \varphi_i(\mathbf{x}(t))] := x_{ij}(t)\delta_{ij}(\mathbf{x}(t)) := f_{ij}(\mathbf{x}(t)), \quad \forall i \in I, j \in N_i \quad (3)$$

其中，$h_{ij}(\mathbf{x}(t))$ 表示第 $i$ 个出行者采取第 $j$ 个纯策略的适应度，它是所有类型出行者纯策略概率分布 $\mathbf{x}(t)$ 的函数，表达式为：

$$h_{ij}(\mathbf{x}(t)) = \sum_{k_1 \in \{1,\cdots,w_1\}} \cdots \sum_{k_n \in \{1,\cdots,w_n\}} x_{1,k_1}(t) \cdots x_{i-1,k_{i-1}}(t) x_{i+1,k_{i+1}}(t) \cdots x_{n,k_n}(t) u_i(\mathbf{e}_{1,k_1}, \cdots, \mathbf{e}_{i-1,k_{i-1}}, \mathbf{e}_{i,j}, \mathbf{e}_{i+1,k_{i+1}}, \cdots, \mathbf{e}_{n,k_n})$$

(4)

$t$ 时刻第 $i$ 个出行者的平均支付

$$\varphi_i(\mathbf{x}(t)) = \sum_{k_1 \in \{1,\cdots,w_1\}} \cdots \sum_{k_n \in \{1,\cdots,w_n\}} x_{1,k_1}(t) \cdots x_{i,k_i}(t) \cdots x_{n,k_n}(t) u_i(\mathbf{e}_{1,k_1},\cdots,\mathbf{e}_{i,k_i},\cdots,\mathbf{e}_{n,k_n}) \quad (5)$$

注意到函数 $h_{ij}(\mathbf{x}(t))$ 的最高次数为 $n-1$ 次，而函数 $\varphi_i(\mathbf{x}(t))$ 的最高次数为 $n$ 次，根据(3)式的定义可知每个 $f_{ij}(\mathbf{x}(t))$ 的次数为 $n$ 次。对每种类型的出行群体，可以按照(3)~(5)式写出其复制者动态方程，对于任意 $i$，$x_{ij}(t)$ 需要满足概率约束条件，故自由变量个数只有 $w_i - 1$ 个。

记 $w$ 维向量
$$\begin{aligned}\mathbf{y}(t) &= (y_1(t),\cdots,y_{w_1}(t),\cdots,y_{w-w_n+1}(t),\cdots,y_w(t))^T \\ &:= (x_{11}(t),\cdots,x_{1w_1}(t),\cdots,x_{n1}(t),\cdots,x_{nw_n}(t))^T\end{aligned},$$

向量值函数
$$\begin{aligned}g(\mathbf{y}(t)) &= [g_1(\mathbf{y}(t)),\cdots,g_{w_1}(\mathbf{y}(t)),\cdots,g_{w-w_n+1}(\mathbf{y}(t)),\cdots,g_w(\mathbf{y}(t))]^T \\ &:= [f_{11}(\mathbf{x}(t)),\cdots,f_{1w_1}(\mathbf{x}(t)),\cdots,f_{n1}(\mathbf{x}(t)),\cdots,f_{nw_n}(\mathbf{x}(t))]^T\end{aligned}$$

则流量演化系统的动力学模型（Flow evolutionary dynamic model，FEDM）可表示为如下向量形式：

$$\frac{d\mathbf{y}(t)}{dt} = g(\mathbf{y}(t)), \mathbf{y}(t) \in \Omega. \quad (6)$$

$$\Omega = \{(y_1(t),\cdots,y_{w_1}(t),\cdots,y_{w-w_n+1}(t),\cdots,y_w(t))^T : 0 \leq y_k(t) \leq 1, \sum_{k=1}^{w_{j+1}} y_{w_1+w_2+\cdots+w_j+k}(t) = 1, k \in W, j \in J\}$$

记指标集合 $J = \{0,1,\cdots,n-1\}$，令 $w_0 = 0$，则 $\Omega \subset \mathbb{R}^w$ 是 $w$ 维实空间 $\mathbb{R}^w$ 中的一个区域。

由于向量值函数 $g(\mathbf{y}(t))$ 中各分量函数 $f_{ij}(\mathbf{x}(t))$ 是关于 $\mathbf{y}(t)$ 中元素的 $n$ 次实系数多项式，故向量值函数 $g(\mathbf{y}(t))$ 在 $\Omega$ 上连续，即 $g(\mathbf{y}(t)) \in C[\Omega, \mathbb{R}^w]$。

下面对多维联合出行下的流量演化动力学模型(6)解的性质（存在性、唯一性和稳定性）进行分析。首先将证明动态用户最优(DUO)下的流量等价于流量演化动力学模型中的纯策略 Nash 均衡，动态随机用户最优(DSUO)下的流量等价于流量演化动力学模型中的混合策略 Nash 均衡，然后我们利用常微分方程解的存在唯一性定理来判断演化动力学模型解的存在性和唯一性，再使用 Liapunov 直接法讨论演化动力学模型解的稳定性。

## 4. 等价性证明

本小节将分别证明动态用户最优均衡等价于流量演化动力学模型的纯策略 Nash 均衡，动态随机用户最优均衡等价于混合策略 Nash 均衡。这一步的证明将从博弈论的角度对动态交通流分配模型均衡解的含义进行解释，它不仅是交通流动态均衡和流量演化动力学分析二者之间联系的桥梁，而且使本文应用流量演化模型对交通流动态均衡展开研究的合理性得到保证。不失一般性，在本小节的证明中，都是取道路网络中任一 OD 对为研究对象，设 OD 对上总共有 $K$ 条路径。

## 4.1 DUOE 和纯策略 Nash 均衡等价性

这里将路径看作是出行者，用小写字母 $k$ 表示第 $k$ 条路径（第 $k$ 个出行者），对于动态流量分配问题，出行者的策略是确定路径上当前时刻的车辆负荷（车辆数），用 $f_k(t)$ 表示 $t$ 时刻路径 $k$ 上的车辆数，也即第 $k$ 个出行者的策略，记策略向量 $\mathbf{s}(t) = (f_1(t), \cdots, f_K(t))^T = (f_k(t), f_{-k}(t))^T$ 表示 $t$ 时刻 $K$ 个出行者的策略。$u_k(s(t)) = u_k(f_k(t), f_{-k}(t))$ 表示 $t$ 时刻当系统出行者采用策略 $s(t)$ 时，第 $k$ 个出行者的收益，取值为负的瞬时路径旅行阻抗，即：

$$u_k(f_k(t), f_{-k}(t)) = -\tau_k(f_k(t), f_{-k}(t)) \tag{7}$$

这里，$\tau_k(f_k(t), f_{-k}(t)) > 0$ 表示 $t$ 时刻第 $k$ 条路径上的瞬时旅行阻抗，它是 $t$ 时刻整个 OD 对车辆负荷分布 $\{f_1(t), \cdots, f_K(t)\}$ 的函数，$\tau^*(t)$ 表示 $t$ 时刻 OD 对间的最小瞬时阻抗。

在证明等价性以前，这里给出两个前提假设：

**假设 1**：理性人假设。即出行者总是从它所知道的路径旅行阻抗中选择阻抗值最小的路径行动；

**假设 2**：阻抗严格单调递增假设。即路径阻抗是当前时刻车辆数的严格单调递增函数，即：

$$[\tau_k(f_k^1(t), f_{-k}^1(t)) - \tau_k(f_k^2(t), f_{-k}^2(t))](f_k^1(t) - f_k^2(t)) > 0, \forall f_k^1(t) \neq f_k^2(t) \tag{8}$$

当 $f_k^1(t) = f_k^2(t)$ 时，上式左端等于 0。

在证明等价性命题之前，先给出动态用户最优（DUO）条件：

**定义 1**（动态用户最优条件[28]）在任一时刻、任一决策节点上，OD 对上所有被使用的路径上的瞬时阻抗都相等且等于最小瞬时阻抗，而所有未被使用的路径上的瞬时阻抗都不小于这个最小瞬时阻抗，称此时的网络状态满足基于瞬时路径阻抗的动态用户最优（DUO）均衡条件。

表示成等价的互补条件就是：

$$[\tau_k(f_k^*(t), f_{-k}^*(t)) - \tau^*(t)] f_k^*(t) = 0, \forall k \in \{1, \cdots, K\} \tag{9}$$

其中，$f_k^*(t), f_{-k}^*(t)$ 分别表示 $t$ 时刻第 $k$ 条路径和其它 $k-1$ 条路径的最优瞬时车辆数目。

**定义 2** （纯策略 Nash 均衡，Pure Strategy Nash Equilibrium，PSNE）称 $t$ 时刻纯策略 $s^*(t) = (f_k^*(t), f_{-k}^*(t))^T$ 满足纯策略 Nash 均衡条件，若

$$u_k(f_k^*(t), f_{-k}^*(t)) \geq u_k(f_k(t), f_{-k}^*(t)), \forall k \in \{1, \cdots, K\} \tag{10}$$

**引理 1** 动态用户最优均衡满足纯策略 Nash 均衡，即 DUO $\Rightarrow$ PSNE。

证明：$\forall k \in \{1, \cdots, K\}$，若 $f_k^*(t) > 0$，则由 DUO 的定义 1 可知：

$$\tau_k(f_k^*(t), f_{-k}^*(t)) = \tau^*(t) \leq \tau_k(f_k(t), f_{-k}^*(t)) \tag{11}$$

若 $f_k^*(t) = 0$，则由于 $f_k(t) \geq 0, \forall k$，由假设 2 可知：

$$\tau_k(f_k^*(t), f_{-k}^*(t)) = \tau_k(0, f_{-k}^*(t)) \leq \tau_k(f_k(t), f_{-k}^*(t)) \tag{12}$$

综合(11)，(12)及(7)式可知，(10)式成立，故引理 1 成立，证毕。

**引理 2** 纯策略 Nash 均衡满足动态用户最优均衡条件，即 PSNE $\Rightarrow$ DUO。

证明：已知(10)式成立，现要证(9)式成立。由(10)成立及(7)可知，

$$\tau_k(f_k^*(t), f_{-k}^*(t)) \leq \tau_k(f_k(t), f_{-k}^*(t)), \forall k \in \{1, \cdots, K\} \tag{13}$$

(1) 若 $f_k^*(t) = 0$，自然有(9)式成立；

(2) 若 $f_k^*(t) > 0$，但 $\tau_k(f_k^*(t), f_{-k}^*(t)) > \tau^*(t)$，由 $\tau^*(t)$ 的含义可知，必至少存在某条路径 $l \in \{1, \cdots, K\}$，使得 $\tau_l(f_l^*(t), f_{-l}^*(t)) = \tau^*(t)$，则对于两条路径 $k, l$，有 $\tau_k(f_k^*(t), f_{-k}^*(t)) > \tau_l(f_l^*(t), f_{-l}^*(t))$，根据假设条件(1)，在理性人假设条件下，必然会有出行者从路径 $k$ 转移到路径 $l$ 上，不妨设转移出的车辆数目为 $\Delta f_{k \to l}(t) > 0$，则路径 $k$ 转移后的新车辆数为 $f_k(t) = f_k^*(t) - \Delta f_{k \to l}(t) < f_k^*(t)$，由路径阻抗的严格单调性假设条件(2)可知，$t$ 时刻路径 $k$ 的阻抗满足 $\tau_k(f_k^*(t), f_{-k}^*(t)) > \tau_k(f_k(t), f_{-k}^*(t))$，这与(13)矛盾。故若 $f_k^*(t) > 0$，必有 $\tau_k(f_k^*(t), f_{-k}^*(t)) = \tau^*(t)$ 成立，综合(1)、(2)两种情况可知，当(10)得到满足时，(9)式成立，故引理 2 成立，证毕。

**推论 1** 动态用户最优均衡等价于纯策略 Nash 均衡，即 DUO $\Leftrightarrow$ PSNE。

证明：由引理 1 和引理 2 即得。

### 4.2 DSUO 和混合策略 Nash 均衡等价性

假设网络中出行者集合 $I = \{1, \cdots, n\}$，这里之所以采用与 4.1 小节中不同的博弈要素设置，是由问题本身的性质所决定的，因为动态用户最优均衡条件中需要确定的是当前时刻路径上的车辆数，而动态随机出行者最优均衡条件需要确定的是出行者选择每条路径的概率，所以在 4.1 小节的证明中，将路径看作博弈参与方，将路径上的车辆数作为策略，而在本小节的证明中，将出行者看作博弈参与方，将出行者选择路径的概率作为策略。

设 $t$ 时刻出行者 $i$ 的策略向量 $\mathbf{s}_i(t) = (x_{i,1}(t), \cdots, x_{i,K}(t))$，这里 $\mathbf{s}_i(t)$ 中的每一个分量满足

概率条件 $0 \leq x_{i,j}(t) \leq 1, \sum_{j=1}^{K} x_{i,j}(t) = 1$，$t$ 时刻第 $i$ 个出行者的收益向量函数为：

$$\mathbf{u}_i(\mathbf{s}_i(t), \mathbf{s}_{-i}(t)) = [u_{i,1}(\mathbf{s}_i(t), \mathbf{s}_{-i}(t)), \cdots, u_{i,K}(\mathbf{s}_i(t), \mathbf{s}_{-i}(t))]^T \tag{14}$$

这里，$\mathbf{s}_{-i}(t) = (\mathbf{s}_1(t), \cdots, \mathbf{s}_{i-1}(t), \mathbf{s}_{i+1}(t), \cdots, \mathbf{s}_n(t))^T$ 表示 $t$ 时刻其它 $n-1$ 个出行者的策略，(14)式中每个分量 $u_{i,k}(\mathbf{s}_i(t), \mathbf{s}_{-i}(t))$ 表示 $t$ 时刻第 $i$ 个出行者采取策略 $\mathbf{s}_i(t)$，而其它 $n-1$ 个出行者采取策略 $\mathbf{s}_{-i}(t)$ 时，第 $i$ 个出行者在第 $k$ 条路径上可以获得的收益，类似于(7)式，这里同样用负的路径旅行阻抗表示收益值，即：

$$u_{i,k}(\mathbf{s}_i(t), \mathbf{s}_{-i}(t)) = -\tau_{i,k}(\mathbf{s}_i(t), \mathbf{s}_{-i}(t)) \tag{15}$$

在证明等价性之前，先给出动态随机用户最优均衡条件 DSUO 和混合策略 Nash 均衡条件。

**定义 3** （动态随机用户最优均衡条件[28]）在任一时刻，在任一 OD 对之间的每一个决策节点上，若从决策节点至终讫点的路径被选用，则这些被选用路径的当前时刻的理解瞬时阻抗等于从决策节点至终讫点的最小理解瞬时阻抗，称此时的网络流量状态满足动态随机用户最优（DSUO）均衡条件。

所谓理解阻抗为最小，就是在当前时刻，出行者按照一定的概率选择出行路径，并且在这个选择概率下认为自己所选择路径的旅行阻抗为最小，写成公式就是：

$$\tau_{i,k}(\mathbf{s}_i^*(t), \mathbf{s}_{-i}^*(t)) \leq \tau_{i,k}(\mathbf{s}_i(t), \mathbf{s}_{-i}^*(t)), \forall i \in I, k \in \{1, \cdots, K\} \tag{16}$$

这里，$\mathbf{s}^*(t) = (\mathbf{s}_i^*(t), \mathbf{s}_{-i}^*(t))^T$ 表示系统处于 DSUO 状态时网络中所有个体出行者的出行策略。

**定义 4** （混合策略 Nash 均衡，Mixed Strategy Nash Equilibrium，MSNE）称 $t$ 时刻混合策略 $\mathbf{s}(t) = (\mathbf{s}_i(t), \mathbf{s}_{-i}(t))^T$ 满足混合策略 Nash 均衡条件，如果 $t$ 时刻第 $i$ 个出行者的收益满足：

$$\mathbf{u}_i(\mathbf{s}_i^*(t), \mathbf{s}_{-i}^*(t)) \geq \mathbf{u}_i(\mathbf{s}_i(t), \mathbf{s}_{-i}^*(t)), \forall i \in I \tag{17}$$

**引理 3** 动态随机用户最优均衡满足混合策略 Nash 均衡，即 DSUO $\Rightarrow$ MSNE。

证明：由于 DSUO 被满足，由定义 3 可知，(16)式成立，再由(14)及(15)式可知，(17)式成立，故引理 3 成立，证毕。

**引理 4** 混合策略 Nash 均衡满足动态随机用户最优均衡，即 MSNE $\Rightarrow$ DSUO。

证明：由 (17) 式成立及 (14) 式可知，对于每一条路径 $k$，有 $u_{i,k}(\mathbf{s}_i^*(t), \mathbf{s}_{-i}^*(t)) \geq u_{i,k}(\mathbf{s}_i(t), \mathbf{s}_{-i}^*(t)), \forall i \in I$ 成立，又由(15)式即得(16)式成立，故引理 4 成立，证毕。

**推论 2** DSUO 解等价于混合策略 Nash 均衡，即 DSUO $\Leftrightarrow$ MSNE。

证明：由引理 3 和引理 4 即得。

由推论 1 和推论 2 可知，当流量系统处于演化稳定状态（即达到流量演化动力学模型的平衡点）时，此时道路网络上的流量处于动态（随机）用户最优均衡状态，下面讨论这种最优均衡状态的稳定性和平衡点附近的长期演化趋势。

**5. 流量演化系统动力学模型解的性质分析**

现在对联合出行选择情形下的流量演化系统的动力学模型 FEDM 的解的性质进行分析。关于流量演化系统 FEDM 的解的存在性和唯一性有如下命题：

**命题**（FEDM 解的存在唯一性） $\forall (t_0, \mathbf{y}_0) \in T \times \Omega$，总存在常数 $t^* > 0$，在区间 $\hat{D} = [t_0 - t^*, t_0 + t^*]$ 内流量演化系统 FEDM 存在唯一的解 $\mathbf{y}(t; t_0, \mathbf{y}_0)$，这里 $T = [0, +\infty) \subset \mathbb{R}$。

考虑到篇幅原因，这里仅给出一个证明思路。因为 $g(\mathbf{y}(t)) \in C[\Omega, \mathbb{R}^w]$，所以只需要看 $g(\mathbf{y}(t))$ 是否在区域 $\Omega$ 上满足 Lipschitz 条件，我们考察 $g(\mathbf{y}(t))$ 对 $\mathbf{y}(t)$ 中任一分量 $\mathbf{y}_l(t)$ 的偏导数的范围。根据 $\mathbf{y}(t)$ 和 $g(\mathbf{y}(t))$ 的定义可知，$g_k(\mathbf{y}(t))$ 与 $f_{ij}(\mathbf{x}(t))$，$y_l(t)$ 与 $x_{pq}(t)$ 存在一一对应，故

$$\frac{\partial g_k(\mathbf{y}(t))}{\partial y_l(t)} = \frac{\partial f_{ij}(\mathbf{x}(t))}{\partial x_{pq}(t)}, \quad \forall i, p \in I; j \in N_i; q \in N_p; k, l \in W。 \tag{18}$$

通过分析函数 $f_{ij}(\mathbf{x}(t))$ 的各项特征可知，$\frac{\partial f_{ij}(\mathbf{x})}{\partial x_{pq}}$ 存在上界，则 $\frac{\partial g_k(\mathbf{y}(t))}{\partial y_l(t)}$ 亦存在上界，故 $g(\mathbf{y}(t)) = [g_1(\mathbf{y}(t)), \cdots, g_w(\mathbf{y}(t))]^T$ 在 $\Omega$ 上满足 Lipschitz 条件，又 $g(\mathbf{y}(t)) \in C[\Omega, \mathbb{R}^w]$，故由常微分方程解的存在唯一性定理可知，$\forall (t_0, \mathbf{y}_0) \in T \times \Omega$，总存在常数 $t^* > 0$，使得在区间 $\hat{D} = [t_0 - t^*, t_0 + t^*]$ 内存在唯一的解 $\mathbf{y}(t; t_0, \mathbf{y}_0)$ 满足流量演化系统 FEDM。

下面简要介绍流量演化动力学模型 FEDM 在平衡点附近稳定性的分析思路。因为 $g(\mathbf{y} = 0) = 0$，所以 FEDM 存在零解，只需要讨论 FEDM 在零解的稳定性即可，其基本思路是根据 Lyapunov 直接法给出它在零解稳定性的充分条件。令 Lyapunov 函数 $V(\mathbf{y}) = \sum_{k=1}^{w} y_k^2$，则函数 $V(\mathbf{y})$ 关于 FEDM 对 $t$ 的全导数为：

$$\frac{dV(\mathbf{y})}{dt} = \sum_{k=1}^{w} \frac{\partial V}{\partial y_k} \cdot g_k(\mathbf{y}) = 2 \sum_{k=1}^{w} y_k^2 \delta_k(\mathbf{y}), \tag{19}$$

由 $g_k(\mathbf{y})$ 与 $f_{ij}(\mathbf{x})$，$y_k$ 与 $x_{ij}(t)$ 存在的对应关系可知，$g_k(\mathbf{y}) = y_k[h_k(\mathbf{y}) - \varphi_k(\mathbf{y})] := y_k \delta_k(\mathbf{y})$，

$\forall k \in W$，因此（19）式第 2 个等式成立。解出使得 $\frac{dV(\mathbf{y})}{dt}$ 大于零，等于零和小于零的 $\mathbf{y}$ 所在的区域，通过 Lyapunov 稳定性和切塔耶夫不稳定性定理，可以得到流量演化动力学模型 FEDM 在零解稳定、渐进稳定和不稳定成立的充分条件。

综合上面关于流量演化动力学模型 FEDM 解的性质的分析可知，如果我们已经建立了交通流的演化动力学模型，那么对于任意可行的初始状态，存在唯一的依时间变化的出行策略满足该系统的演化动力学，而且，可以进一步判断该演化系统的平衡点关于初始值的稳定性质，这里关键是需要确定各个出行者在出行过程中各方的收益（支付）参数，结合收益参数的大小来判断上述稳定性判定的三个条件哪一个得到满足，从而预测系统流量的长期演化趋势。

**6、数值算例**

本小节用一个简单的数值算例说明本文建立的流量演化动力学模型的合理性和有效性。设路网 $G(N,A)$ 由两节点、两条路段组成，节点集 $N=\{O,D\}$，路段集 $A=\{1,2\}$，路网拓扑结构如下图 2 所示：

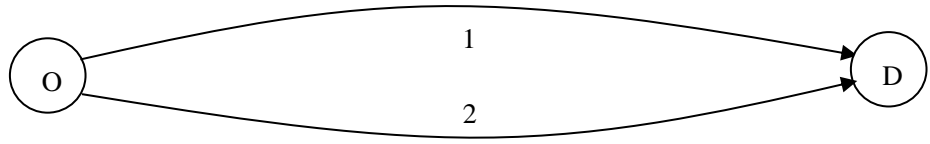

图 2  算例路网拓扑结构图

Figure 2  Schematic diagram of example networks

路网参数：$l_1=3\,km$，$l_2=2\,km$，$v_1=30\,m/s$，$v_2=20\,m/s$，$k_{j,1}=0.25\,veh/m$，$k_{j,2}=0.25\,veh/m$，$l_1$ 和 $l_2$ 分别为路段 1、2 长度，$v_1$ 和 $v_2$ 分别为路段 1、2 上的自由流速度，$k_{j,1}$ 和 $k_{j,2}$ 分别是路段 1 和路段 2 的阻塞密度，路段特性函数[29]：

$$c_a(k_a(t))=\frac{l_a k_j}{v_{\max}(k_j-k_a(t))}. \qquad (20)$$

其中，$k_a(t)\equiv\frac{x_a(t)}{l_a}$ 为 $t$ 时刻路段 $a$ 上的交通密度，单位 veh/m；$v_{\max}$ 为自由流速度，m/s；$k_j$ 为阻塞密度，单位：veh/m。路段 1、路段 2 上的车辆负荷分别为 $x_1(t)$，$x_2(t)$，为了保证 $c_a(k_a(t))>0$，必须有路段负荷 $x_1(t)<750$，$x_2(t)<500$。

**6.1 动态用户最优（DUO）模型**

设所考虑的建模时间段为 $[0,T]$，建立满足动态用户最优（DUO）均衡的分配模型[28, 29]如下：

$$u_a(t)(c_a(t)-c_{\min}^{OD})=0;\ a\in A. \qquad (21)$$

$$\int_0^T \sum_{a \in A} (u_a(t) - u_a^*(t))(c_a(t) - c_a^*(t))dt \geq 0 \ . \tag{22}$$

$$\frac{dx_a(t)}{dt} = u_a(t) - v_a(t); a \in A \ . \tag{23}$$

$$\frac{dc_a(t)}{dt} = \frac{u_a(t)}{v_a(t + c_a(t))} - 1; a \in A \ . \tag{24}$$

$$\sum_{a \in A} x_a(t) = D(t) \ . \tag{25}$$

$$x_a(t) \geq 0, u_a(t) \geq 0, v_a(t) \geq 0 \ . \tag{26}$$

符号说明：

$c_a(t)$：$t$ 时刻路段 $a$ 上的实际行驶阻抗；

$c_{\min}^{OD}$：OD 对间的最小行驶阻抗；

$x_a(t)$：$t$ 时刻路段 $a$ 的交通负荷（车辆数），单位：veh；

$u_a(t)$ 和 $v_a(t)$ 分别为 $t$ 时刻路段 $a$ 上的流入率和流出率；

$u_a^*(t)$ 和 $c_a^*(t)$ 分别为 $t$ 时刻，满足 DUO 条件下路段 $a$ 上的流入率和行驶阻抗；

$D(t)$：$t$ 时刻 OD 对间的交通需求负荷量，这里假设固定为 600，单位：veh；

其中，（21）为动态用户最优（DUO）条件，（22）为等价的变分不等式（VI）问题；（23）为状态约束条件，（24）为流量传播约束条件，（25）为守恒约束，（26）为非负约束。

将路段参数代入（22）～（26），可以解出状态变量 $x_1(t) \approx 436 \approx 0.726D(t)$，

$x_2(t) \approx 164 \approx 0.274D(t)$。

## 6.2 流量演化动力学模型（FEDM）

出行者 {A,B}，策略集 {1,2}，2×2 对称博弈的规范形式：$\begin{pmatrix} A,B & 1 & 2 \\ 1 & (a,a) & (b,c) \\ 2 & (c,b) & (d,d) \end{pmatrix}$，

其中，博弈支付参数 $a$ 表示当出行者 A 选择路段 1，对手 B 也选择路段 1，A 获得的收益，这里收益用负的阻抗来表示，其它字母 $b$、$c$、$d$ 的含义类似，A 的博弈支付矩阵 $\begin{pmatrix} a & b \\ c & d \end{pmatrix}$，

在动态交通环境下，由公式（26）可知，道路行驶阻抗函数是路段交通负荷的严格单调递增函数，则易知博弈支付参数满足：

$$a < c, \ b > d \ . \tag{27}$$

在博弈策略分别为 {1,1}，{1,2}，{2,1}，{2,2} 时，路段 1 和 2 上面的交通负荷量分别为

$\{x_1(t)+x_2(t)=D(t),0\}$，$\{x_1(t),x_2(t)\}$，$\{x_2(t),x_1(t)\}$，$\{0,x_1(t)+x_2(t)=D(t)\}$，将交通负荷量分别代入公式（26）中，可得支付参数如下：

$$a=\frac{100}{\frac{x_1(t)+x_2(t)}{750}-1}, \quad b=\frac{100}{\frac{x_1(t)}{750}-1}, \quad c=\frac{100}{\frac{x_1(t)}{500}-1}, \quad d=\frac{100}{\frac{x_1(t)+x_2(t)}{500}-1}. \tag{28}$$

由路段负荷条件 $x_1(t)<750, x_2(t)<500$ 易知 $a,b,c,d$ 均为负数。

由（27）和（28）可得，$x_1$ 和 $x_2$ 满足约束条件 $x_1<2x_2$。

设总需求 $D(t)=600$ 中，路段 1 上面分配的交通负荷比例为 $p_1=\dfrac{x_1(t)}{x_1(t)+x_2(t)}$，路段 2 上面分配的交通负荷比例为 $p_2=1-p_1=\dfrac{x_2(t)}{x_1(t)+x_2(t)}$，由 $x_1<2x_2$ 易知 $p_1<\dfrac{2}{3}, p_2>\dfrac{1}{3}$，可得流量演化动力学模型 FEDM[30]：

$$\dot{p}_1 = p_1(\mathbf{e}_1^T - \mathbf{p}^T)A\mathbf{p} := f(p_1). \tag{29}$$

其中，$\mathbf{e}_1=(1,0)^T$，$\mathbf{p}=(p_1,p_2)^T$。

### 6.3 结果分析和讨论

对流量演化动力学模型（29）化简得

$$\dot{p}_1 = f(p_1) := p_1(1-p_1)((a-c+d-b)p_1+b-d) \\ = (a-c+d-b)p_1(1-p_1)(p_1-p^*) \tag{30}$$

，这里，$p^*=\dfrac{d-b}{a-c+d-b}$. \quad (31)

易知，流量演化系统（29）存在唯一演化稳定策略（ESS）：$\{p_1^*=p^*\}$。

（1）演化动力学模型 FEDM 解的存在唯一性

将 $f(p_1)$ 对 $p_1$ 求导数取绝对值得：

$$\left|\frac{df(p_1)}{dp_1}\right| = |(1-p_1)(a'p_1+b')-p_1(a'p_1+b')+a'p_1(1-p_1)| \\ \leq 3|a'|+2|b'|=3(c-a)+5(b-d)$$

这里，$a'=a-c+d-b<0$，$b'=b-d>0$，则由于 $f(p_1)\in C^1[0,1]$，在 $[0,1]$ 上满足 Lipschitz 条件，故流量演化系统（29）对于某个初值 $(t_0,p_1)$ 在 $[0,1]$ 内存在唯一解。

（2）FEDM 的演化稳定点（ESS）与动态用户最优（DUO）条件的等价性

将 $x_1(t)=p^*\cdot D(t)$，$x_2(t)=(1-p^*)\cdot D(t)$ 和支付参数(28)代入(31)中，可以解得

$p^* \approx 0.726$，当时间 $t$ 充分大的时候，有 $x_1(t) \to p^* D(t) \approx 0.726 D(t)$，$x_2(t) \to (1-p^*)D(t) \approx 0.274 D(t)$，即流量演化系统（29）的演化稳定点对应的交通负荷就是满足动态用户最优条件下的解，通过分析流量演化动力学模型（29）得到了满足 DUO 条件的均衡解，证明了本文模型的合理性和有效性。

值得注意的是，本文第 3～第 5 小节中的讨论都是基于网络路径的，而第 6 小节给出的算例却是基于网络路段的，这里有两个原因，首先是由于算例路网的简单性（给出的算例只含有一个 OD 对和两条路段），使得这里的路段同时也是路径，第二个更重要的也是一般性的原因，就是基于路径流量和基于路段流量的 DUO 条件之间存在密切的关系，用式子表示就是：$u_a(t) = \sum_{q \in Q_a} f_q(t)$，这里，$u_a(t)$ 为 $t$ 时刻路段 $a$ 的流入率，$f_q(t)$ 为 $t$ 时刻路径 $q$ 的流量速率，$Q_a$ 为包含路段 $a$ 的路径集合。由这个关系式可以推出基于瞬时路径阻抗的 DUO 条件与基于瞬时路段阻抗的 DUO 条件的一致性，即满足基于瞬时路径阻抗的 DUO 条件也满足基于瞬时路段阻抗的 DUO 条件，反之亦然[28]，同时，考虑到基于瞬时路径阻抗的 DUO 条件下的动态流量分配模型在设计算法时需要列举路径，给实际大型交通网络的应用带来不便，因此，在理论分析中往往采用基于路段的 DUO 动态分配模型来代替基于路径的 DUO 分配模型，一个这两者本质上一致，二是会带来求解的便利，本文 6 小节的算例就是基于这个考虑的。

**7、结束语**

本文针对交通流从不均衡到均衡的动态演化过程，提出了交通流演化动力学研究的一般框架，建立了流量演化的微分动力学模型，证明了动力学模型对应的纯策略 Nash 均衡（PSNE）和动态用户最优（DUO）条件以及混合策略 Nash 均衡（MSNE）和动态随机用户最优（DSUO）条件之间的等价性，同时，动力学模型的 ESS 就是满足 DUO 的均衡解，在动态交通流分配问题和出行者博弈行为之间建立了联系，并讨论了动力学模型解的性质（解的存在性、唯一性和平衡点附近的稳定性）。这些结论有助于我们加深对于城市交通流演化规律的认识和理解，对于发展出有效的交通管理控制措施也具有积极的理论价值和实际意义。本文提出的交通流演化动力学的研究框架具有相当的一般性，可以很方便地推广到其他更为复杂的出行行为决策下的流量演化相关问题的研究中。

本文只是从理论分析的角度对交通流的动态均衡和演化动力学问题进行了探讨，考虑到现实问题的复杂性，模型对实际情况进行了大量的假设和简化，下一步可以运用仿真和实证的方法，对本文的相关结论展开验证，另外，对本文中的一些参数具体取值的设定，也是值得进一步深入探讨的问题。

# General Research Framework and its Properties Analysis of Traffic Flow Evolutionary Dynamics

**Abstract：**

This paper aims to the conditions of traffic flow evolving to stability and the stability of equilibrium under demand time-varying of traffic networks. The general framework of the evolution of flow dynamics by adopting evolutionary game theory and dynamic system stability theory is proposed. The multi group and multi criteria travel choice flow evolution dynamic model is carried out. The equivalence of the balance position of flow evolution dynamic model and equilibrium solutions of dynamic traffic assignment model is proved. Furthermore, the existence, uniqueness and stability of the solutions are discussed. Theoretical analysis results show that unique solutions of evolutionary dynamic model are always existent in a local region. Different stability properties will emerge nearby the equilibrium solutions of flow evolution dynamic model under the hypothesis of traveler individual income parameters satisfying certain conditions. Rationality and validity of the proposed model are addressed through a simple numerical example. Links between travelers' game behavior and dynamic traffic assignment is established. These conclusions can help us deepen our understanding to urban traffic flow evolution law.
**Keywrods**: urban traffic; traffic flow; dynamics; traffic assignment, evolutionary game